\documentclass[preprint,aps]{revtex4}

\usepackage{graphicx}% Include figure files
\usepackage{dcolumn}% Align table columns on decimal point
\usepackage{bm}% bold math
\usepackage{subfig}
\usepackage{w-greek}
\usepackage{multirow}

\begin{document}

% Title of the article
\title{On the single-particle-reduced entropy of a gated nanowire system in the Coulomb blockade regime}

% Authors
\author{Jos\'e Mar\'ia Castelo}
\author{Klaus Michael Indlekofer}
\email{michael.indlekofer@hs-rm.de}
\affiliation{
RheinMain University of Applied Sciences, FB ING / Institute of Microtechnologies,
Am Br\"uckweg 26, D-65428, R\"usselsheim, Germany}

\author{J\"org Malindretos}
\affiliation{
Georg-August-Universit\"at G\"ottingen, IV. Physikalisches Institut,
D-37077, G\"ottingen, Germany}

\date{May 29, 2013}

\begin{abstract}
 In this paper, the single-particle-reduced entropy of a nanowire field-effect transistor (NWFET) in the Coulomb blockade regime is studied by means of a multi-configurational self-consistent Green's function approach.
Assuming that the many-body statistical preparation of the system is described by a mixture of Slater determinants of relevant natural orbitals, the single-particle-reduced entropy can be interpreted as a measure of the degree of mixture of the system's preparation.
Considering the realistic case of an InP based NWFET, we present current--voltage characteristics and entropy diagrams for a range of equilibrium and non-equilibrium states. Signatures of few-electron Coulomb charging effects can be identified, as known from experimental situations. Furthermore, we illustrate the significance of the single-particle-reduced entropy by analyzing the corresponding electronic configurations.
\end{abstract}

\pacs{} \keywords{Single-particle-reduced entropy, Coulomb blockade, semiconductor nanowire} \maketitle

% The class file requires the standard graphicx Latex package. See the 'LaTeX
% standard graphics and color packages documentation' for more information at
% <http://tug.ctan.org/tex-archive/macros/latex/required/graphics/grfguide.pdf>.
%
% Accepted figure file formats depend on which LaTeX flavour is used.
% Classic LaTeX is always able to use Encapsulated Postscript (EPS);
% PDFLaTeX can't use this but accepts PDF, JPG, PNG, and GIF formats.
%
% See examples for implementing graphics in floating figure environments later in this file.
% If \titlefigure is given, it takes as its mandatory parameter the
% name (without extension) of some figure file.
%\titlefigure[height=3.1cm]{empty2w}
%\titlefigurecaption{%
%  This is the caption of the \emph{optional} abstract figure. If
%  there is no abstract figure here, the abstract text should be divided into both columns.}

\section{Introduction}
\label{sec:introduction}
Semiconductor nanowires have attracted great interest as candidates for nanoelectronic applications and might become a basis for new and high-performance devices on the nanometer scale \cite{duan2001,thelander2006}. They enable ultimately scaled conventional FETs as well as novel transistor concepts, which are expected to offer superior performence in terms of current drive capability and subthreshold slope \cite{appenzeller2008}. Moreover, nanowire based transistor devices allow for the investigation of fundamental questions of electronic transport in quasi one-dimensional quantum systems.

A manifold of numerical studies exists that approaches the subject of electronic transport in NWFETs using quantum kinetic methods for application-relevant conditions, based on a non-equilibrium Green's function (NEGF) formalism
\cite{lake1999,datta2000,jin2006,shin2007,wang2004,martinez2007}. However, one of the challenges in the realistic simulation of nanodevices consists in the large number of degrees of freedom (e.g. localized orbitals) in combination with many-body Coulomb interaction effects, which make a full Fock-space description numerically unfeasible. Therefore, the majority of nanodevice simulation techniques on realistic length scales are based on a mean-field approximation of the Coulomb interaction. In this context, the multi-configurational Green's function approach \cite{indlekofer2005,indlekofer2007} combines the numerical advantages of a mean-field approximation with a Fock-space description of a small sub-set of relevant, resonantly trapped states which are responsible for Coulomb charging effects due to individual electrons. In this paper, we analyze the role of these relevant states and the resulting configurations by means of the single-particle-reduced entropy \cite{esquivel1996,indlekofer2013} for the case of a realistic InP based nanowire system.

The structure of the paper is as follows. First, we briefly describe the employed quantum kinetic simulation approach, followed by a definition of the considered nanodevice. Then in section \ref{sec:transport}, we focus on the electronic transport in the Coulomb blockade regime and show signatures of few-electron Coulomb charging effects. Section \ref{sec:entropy} is devoted to the analysis of the single-particle-reduced entropy of the system. Finally, the conclusions are drawn.

\begin{figure}[t]
\centering
\includegraphics[width=\linewidth]{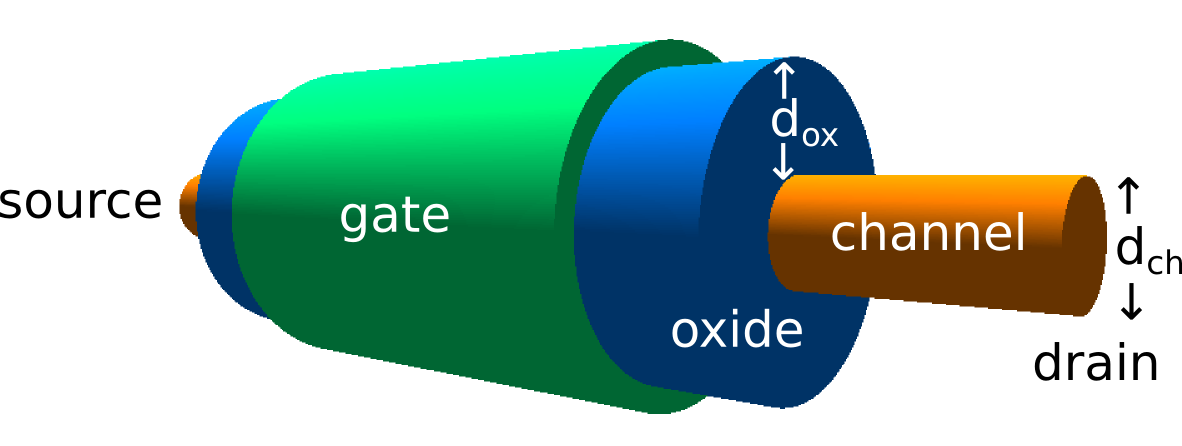}
\caption{Schematic view of the NWFET geometry, where $d_{ox}$ and $d_{ch}$ are the gate insulator thickness and the channel diameter, respectively.}
\label{fig:device}
\end{figure}

\begin{figure*}[t]%
\subfloat[Absolute value of the drain current $|I_D|~\mathrm{(nA)}$.]{\label{fig:diamonds}%
\includegraphics*[width=.5\textwidth]{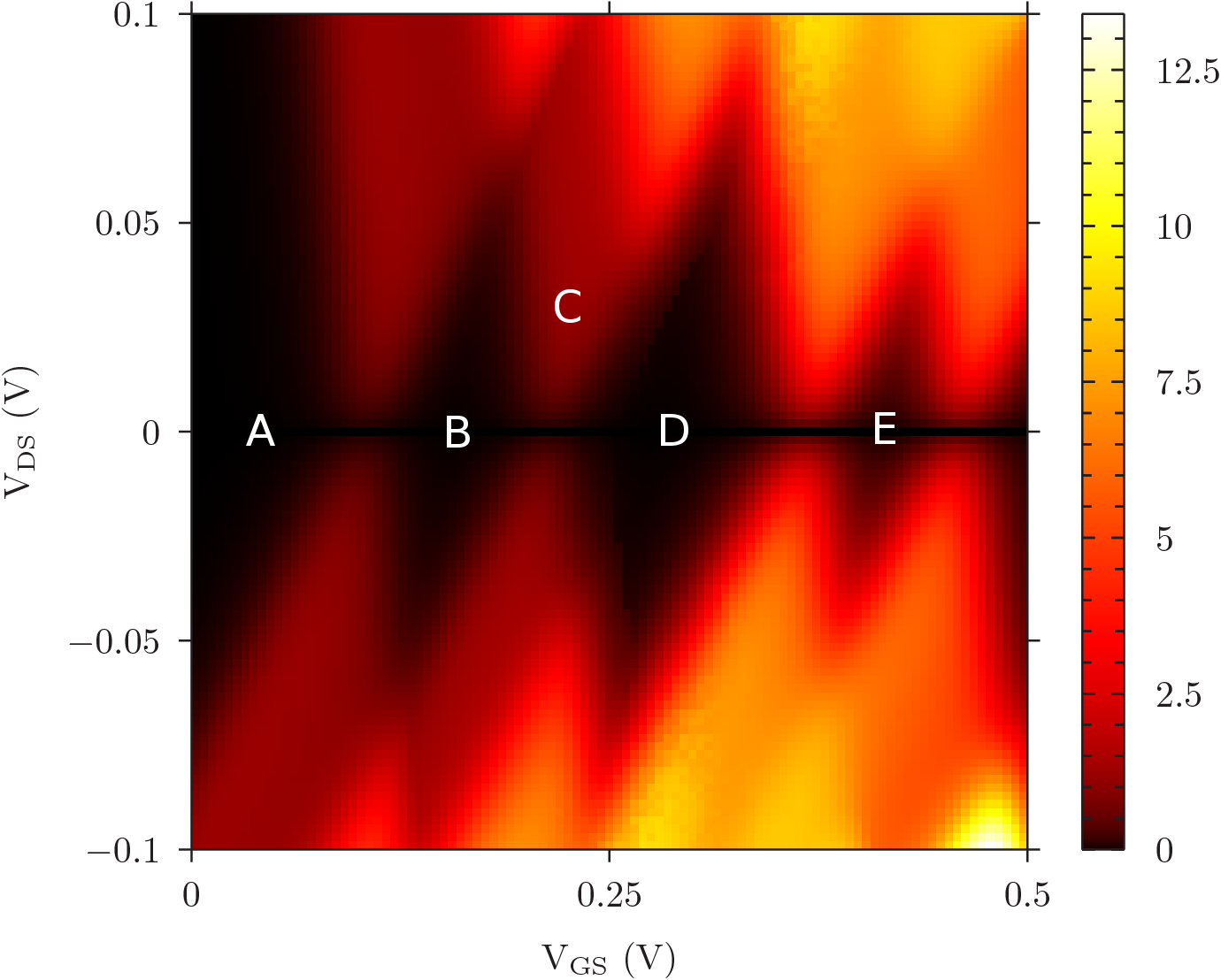}}\hfill
\subfloat[Single-particle-reduced entropy $S_1~\mathrm{(bit)}$.]{\label{fig:entropy}%
\includegraphics*[width=.5\textwidth]{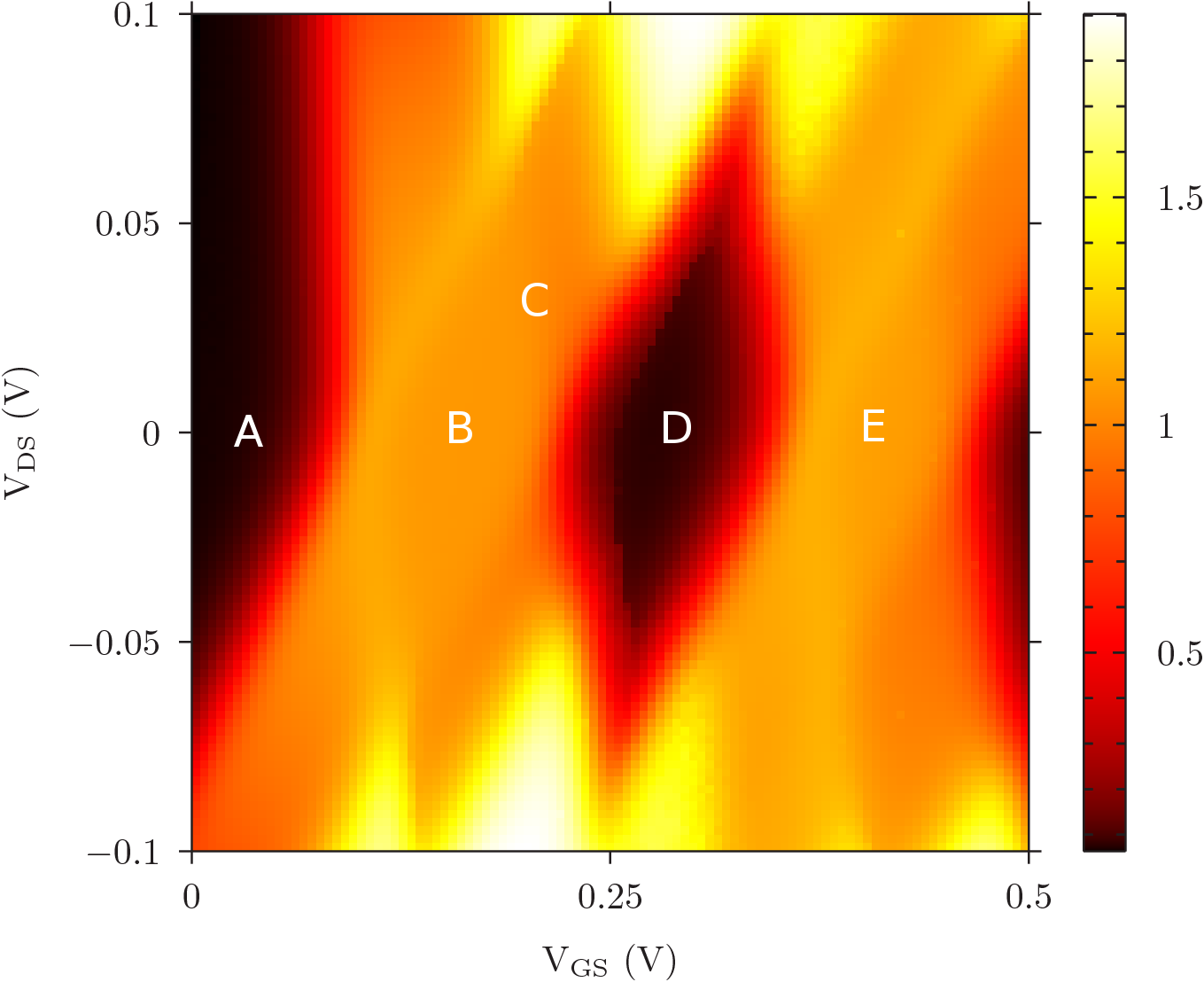}}\hfill
\caption{%
Simulated current--voltage characteristics (a) and single-particle-reduced entropy (b).}
\label{fig:diamonds_and_entropy}
\end{figure*}

\section{Simulation approach}
\label{sec:device}

For our calculations, we have employed a multi-configurational self-consistent Green's function (MCSCG) approach \cite{indlekofer2005,indlekofer2007} for a realistic description of discrete resonant states, beyond the mean-field approximation. We make use of a one-band tight-binding description of the channel in the effective mass approximation, represented by a basis of $2 N_{sites}$ localized 1D single-particle spin orbitals. The benefit of the MCSCG approach stems from the self-consistent division of the channel system into a small subsystem of resonantly trapped (relevant) states for which a many-body Fock-space treatment becomes numerically feasible, and the rest of the system which is treated adequately on a conventional mean-field level. The Fock-space description allows for the calculation of few-electron Coulomb charging effects beyond mean-field, as manifested in the Coulomb blockade regime.

In the following, $\rho_1$  denotes the single-particle density matrix \cite{lowdin1955} of the system under non-equilibrium conditions. For the NWFET channel, $\rho_1$ is a $2N_{sites}\times2N_{sites}$ matrix and is obtained from the NEGF formalism. The single-particle eigenvectors of $\rho_1$ are called \emph{natural orbitals} \cite{lowdin1955}, its eigenvalues are real and within the interval $[0,1]$. For the case of a single Slater determinant, $\rho_1$ would have only eigenvalues 0 and 1.

The MCSCG approach describes the state of the channel with fluctuating electron number as a weighted mixture of a few relevant many-body states, so-called \emph{configurations}. In the current approximation, these configurations are assumed to be Slater determinants of resonantly trapped natural orbitals \cite{nwfetlab2013}. The corresponding weights are real, restricted to the $[0,1]$ interval, and their sum is unity. Thus, the relevant many-body statistical operator consists of a weighted sum of projectors, restricted in our case to a relevant sub-space of the whole Fock-space.

\section{Device layout}
\label{sec:transport}

Figure \ref{fig:device} shows the elements of the considered coaxially-gated NWFET. The channel consists of an InP nanowire, which has a lattice constant $a_0=5.9~\mathrm{\AA{}}$, an electron effective mass $m^{\ast}/m_e=0.079$ and a relative dielectric constant $\epsilon_r=12.5$ \cite{adachi1992}. We chose a basis with a number of sites $N_{sites}=30$ giving a length $L=17.7~\mathrm{nm}$. Its diameter is $d_{ch}=5~\mathrm{nm}$. The channel and the gate are separated by a $\mathrm{SiO_2}$ oxide layer of thickness $d_{ox}=10~\mathrm{nm}$ and relative dielectric constant $\epsilon_r=3.9$. The source and drain are contacted with Pd/Ti to form Schottky barriers \cite{reddy2011} of height $\Phi_{SB}=0.7~\mathrm{eV}$, which act as double tunnel barriers to confine the electrons in the nanowire. The same device is used in all the following sections, with a temperature of $T=77~\mathrm{K}$.

\section{Simulated electronic transport}
\label{sec:transport}

At low enough temperatures and drain--source voltages $V_{DS}$ (compared to the channel's Coulomb charging energy and single-particle energy level spacing), the energy which is necessary to add an extra electron to the channel can exceed the thermal energy, and the current through the NWFET is blocked. Within this Coulomb blockade regime, the NWFET behaves as a single-electron transistor (SET) \cite{kastner1992,houten1992,kouwenhoven2001}.

We have simulated electronic transport in this regime by means of the MCSCG approach. Coulomb diamonds emerge in the current--voltage characteristics, as shown in Fig.~\ref{fig:diamonds}. In these regions, the drain current $I_D$ reduces to almost zero by Coulomb blockade and the channel is occupied by an integer number of electrons $N_e$. Along the $V_{DS}\simeq0$ axis, drain current can flow where the electron number within the channel fluctuates between $N_e$ and $N_e+1$. By varying the gate--source voltage $V_{GS}$, Coulomb oscillations can be observed. The simulations correctly resemble the features that are observed in experiments \cite{franceschi2003}.

\begin{table*}[t]
 \centering
 \begin{tabular}{l c c c c c c}
  \hline
  Point & $N_e$ & Shell & $(V_{GS},V_{DS})$ & $S_1$ & Slater determinants & Weights \\
  \hline \hline
  A: Empty channel & 0 & Empty & (0.034, 0) & 0.074 & (00000000) & 1 \\ \hline
  \multirow{2}{*}{B: 1st Diamond} & \multirow{2}{*}{1} & \multirow{2}{*}{Open} & \multirow{2}{*}{(0.15, 0)} & \multirow{2}{*}{1.07}
   & (10000000) & 0.502 \\
   & & & & & (01000000) & 0.498 \\ \hline
  \multicolumn{3}{l}{\multirow{3}{*}{C: Non-equilibrium state}} & \multirow{3}{*}{(0.23, 0.026)} & \multirow{3}{*}{0.908}
   & (10000000) & 0.363 \\
   & & & & & (01000000) & 0.361 \\
   & & & & & (11000000) & 0.276 \\ \hline
   D: 2nd Diamond & 2 & Closed & (0.285, 0) & 0.079 & (11000000) & 1 \\ \hline
  \multirow{3}{*}{E: 3rd Diamond} & \multirow{3}{*}{3} & \multirow{3}{*}{Open} & \multirow{3}{*}{(0.405, 0)} & \multirow{3}{*}{1.102}
   & (11000000) & 0.024 \\
   & & & & & (11100000) & 0.489 \\
   & & & & & (11010000) & 0.487 \\
  \hline
 \end{tabular}
 \caption{Data from selected points in the diagrams. The table shows the electron number $N_e$, shell filling, voltage coordinates $(V_{GS},V_{DS})$, single-particle-reduced entropy $S_1$, the dominant Slater determinants and their associated weights $\{w_i\}$.}
 \label{tab:data}
\end{table*}

\section{Single-particle-reduced entropy}
\label{sec:entropy}

The single-particle-reduced entropy in bit is defined as \cite{esquivel1996,indlekofer2013}
\begin{equation}
 S_1\equiv-\mathrm{Tr}(\rho_1\log_2\rho_1).
 \label{eq:entropy} 
\end{equation}
In quantum chemistry, $S_1$ is referred to as the \emph{correlation entropy} \cite{esquivel1996,gersdorf1997} for pure many-body states.  In general, $S_1$ indicates the deviation from a single Slater determinant, due to correlation and/or mixture. Expanded in terms of the natural orbital basis, $S_1$ can be written as $S_1=-\sum_i n_i \log_2 n_i$, where $n_i$ denote the eigenvalues of $\rho_1$. In this paper, the NWFET channel is described by a mixture of a few relevant Slater determinants with detailed Coulomb interaction, whereas the "non-relevant" rest is treated on a mean-field level. Then, $S_1$ measures the degree of mixture of the system's preparation.

In the extreme case of a many-body preparation that is a mixture of Slater determinants with a set of fully disjoint single-particle states (i.e., with complementary occupation), $N_{SD}=2^{S_1/N_e^f}$ can be interpreted as the number of relevant Slater determinants in the preparation \cite{indlekofer2013}, where $N_e^f$ denotes the number of \emph{fluctuating} electrons. For example, point B in Table~\ref{tab:data} corresponds to $N_e^f=1$.

Figure~\ref{fig:diamonds_and_entropy} shows the simulated characteristics which clearly exhibit a correspondence between the Coulomb diamonds in the current--voltage characteristics and certain diamond-like shaped structures in the entropy diagram. For those regions where the current is blocked and the channel occupied by $0, 1, 2, 3\ldots$ electrons (Coulomb diamonds from left to right) the state of the device jumps consecutively from open shell ($N_e=1,3\ldots$ and multiple Slater determinants) to closed shell ($N_e=2,4\ldots$ and a single Slater determinant). $S_1$ indeed reflects these mixtures and therefore its geometry is also diamond-shaped. This will be seen in detail in the following.

Table~\ref{tab:data} gathers information obtained by means of the MCSCG formalism, corresponding to different voltage points in Fig.~\ref{fig:diamonds_and_entropy}. Point A is associated with an empty channel ($N_e=0$), whose state is given by the vacuum state with unity weight. $S_1\simeq0$ in this case, indicating that the degree of mixture of the preparation is minimum.

Point B corresponds to the first diamond ($N_e=1$). Here, the single electron in the channel ($N_e^f=1$) has the chance to occupy one of the two states in the first single-particle energy level: with spin up or with spin down (open shell configuration). This is reflected in the two Slater determinants ($N_{SD}\simeq2$) with almost equal weights $w_1\simeq w_2\simeq 0.5$ that describe the state of the system. $S_1\simeq1$ and therefore the degree of mixture is higher than in the empty channel case.

A non-equilibrium state is described in point C by three Slater determinants with similar weights $w_1\simeq w_2\simeq w_3\simeq 0.3$. This situation shows that it is equally probable in this state to find an electron in the first single-particle energy level with spin up or down, or two electrons with both spin directions.

Point D corresponds to the second diamond ($N_e=2$). Each electron occupies the first single-particle energy level with spin up and down respectively ($N_e^f=0$). Therefore, there is only one Slater determinant to describe this situation and so its weight is unity. $S_1\simeq0$ as in the case of an empty channel. Again, the degree of mixture is minimum.

Finally, point E shows the case of the third diamond ($N_e=3$), very similar to that of the first diamond. Here, two electrons occupy the first single-particle energy level, and a third one ($N_e^f=1$) has both possibilities of spin occupation in the second single-particle energy level. Although there is a small contribution from a doubly occupied Slater determinant, we can see that $N_{SD}\simeq2$, and the last two determinants are the main contributors ($w_1\simeq0$, $w_2\simeq w_3 \simeq 0.5$). The fact that $S_1$ is slightly higher than one reflects that the first determinant also plays a minor role.

\begin{figure}[t]
\centering
\includegraphics[width=\linewidth]{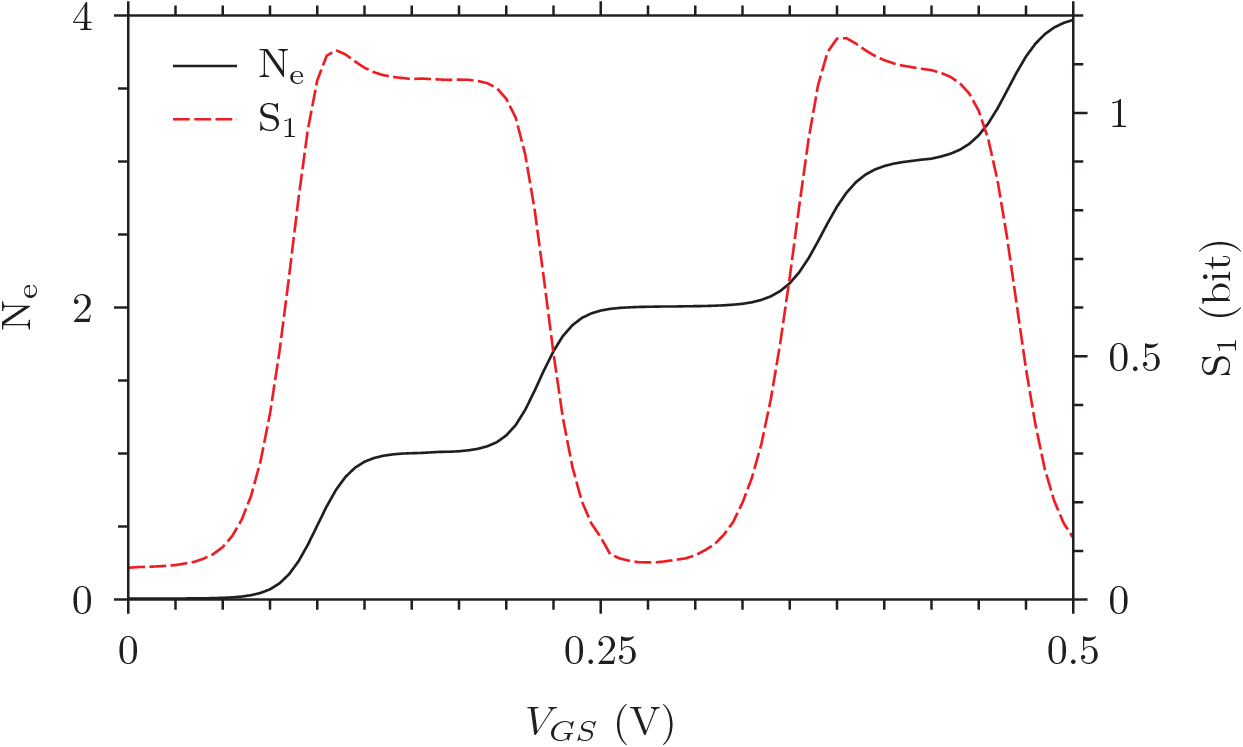}
\caption{Electron number $N_e$ and single-particle-reduced entropy $S_1~\mathrm{(bit)}$ as a function of $V_{GS}$ for fixed $V_{DS}=0$.}
\label{fig:enumber_entropy}
\end{figure}

All of the near-equilibrium cases corresponding to Coulomb diamonds that we have analyzed can be summarized in Fig.~\ref{fig:enumber_entropy}, in which $N_e$ and $S_1$ are plotted against $V_{GS}$ for fixed $V_{DS}=0$. The $N_e$ curve exhibits integer charging steps. On the other hand, $S_1$ oscillates approximately between zero and one. It can be seen here that $S_1\simeq0$ for $N_e \simeq 0 ~\mathrm{and}~ 2$, corresponding to an empty channel and the second Coulomb diamond (closed shell configuration) respectively, described by only one Slater determinant, as shown in Table~\ref{tab:data}. Whereas $S_1\simeq1$ for $N_e \simeq 1 ~\mathrm{and}~ 3$, corresponding to the first and third diamonds (open shell configurations) with more than one Slater determinant. One has to note that the rest of the system (outside the relevant Fock sub-space) also contributes to $S_1$, leading to slightly increased values.

\section{Conclusion}
We have simulated quantum transport properties of a realistic gated nanowire system in the Coulomb blockade regime by means of a multi-configurational Green's function approach. The current--voltage characteristics correctly resemble the experimentally known signatures of few-electron Coulomb charging effects. In addition, the single-particle-reduced entropy has been analyzed for several channel states, showing that it is a measure of the mixture in the system's thermodynamical preparation. In this sense, $S_1$ can be interpreted as the lack of information about the Slater determinant in which the system can be found.

\section{Acknowledgement}
The research leading to these results has received funding from  the European Union Seventh Framework Programme under agreement no. 265073 (nanowiring).

% Use the following code if you wish to generate your bibliography with BibTeX;
% replace the string "pss-demo" below with the name(s) of
% the BibTeX data base(s) you want to use.
% The resulting bibliography-output (the content of the .bbl file)
% must be pasted back into this file before submission.
% Please also include your BibTeX data base file(s) in your submission
% so that we can re-run BibTeX if necessary.
%
%\bibliographystyle{pss}
%\bibliography{pss-demo}
%
% Replace the following example bibliography with your references
% before submission:

\end{document}